\documentstyle[prd,aps,preprint,tighten,epsfig]{revtex}

\begin{document}

\draft

\title{Friedberg-Lee symmetry and tri-bimaximal neutrino mixing \\
in the inverse seesaw mechanism}
\author{{\bf Aik Hui Chan}$^{\rm a,b}$, ~ {\bf Hwee Boon Low}$^{\rm a}$,
~ {\bf Zhi-zhong Xing}$^{\rm c,}$\thanks{E-mail:
xingzz@ihep.ac.cn}}
\address{$^{\rm a}$Institute of Advanced Studies, Nanyang Technological
University, Singapore 639673 \\
$^{\rm b}$Department of Physics, National University of
Singapore, Singapore 117542 \\
$^{\rm c}$Institute of High Energy Physics and Theoretical Physics
Center for Science Facilities, \\ Chinese Academy of Sciences,
Beijing 100049, China\thanks{Permanent address}} \maketitle

\begin{abstract}
The inverse seesaw mechanism with three pairs of gauge-singlet
neutrinos offers a natural interpretation of the tiny masses of
three active neutrinos at the TeV scale. We combine this picture
with the newly-proposed Friedberg-Lee (FL) symmetry in order to
understand the observed pattern of neutrino mixing. We show that
the FL symmetry requires only two pairs of the gauge-singlet
neutrinos to be massive, implying that one active neutrino must be
massless. We propose a phenomenological ansatz with broken FL
symmetry and exact $\mu$-$\tau$ symmetry in the gauge-singlet
neutrino sector and obtain the tri-bimaximal neutrino mixing
pattern by means of the inverse seesaw relation. We demonstrate
that non-unitary corrections to this result are possible to reach
the percent level and a soft breaking of $\mu$-$\tau$ symmetry can
give rise to CP violation in such a TeV-scale seesaw scenario.
\end{abstract}


\newpage

\section{Introduction}

The fact that three active neutrinos possess non-degenerate but
tiny masses is a striking signature of new physics beyond the
standard model \cite{PDG08}. Perhaps the most popular approach
towards understanding the small neutrino mass scale ($< 1$ eV) is
the seesaw mechanism \cite{SS1}, which contains three right-handed
neutrinos and retains $SU(2)^{}_{\rm L} \times U(1)^{}_{\rm Y}$
gauge symmetry. Although this mechanism can naturally work at a
superhigh-energy scale ($\sim 10^{14}$ GeV) to generate tiny
Majorana neutrino masses, it loses direct testability on the
experimental side and causes a hierarchy problem on the
theoretical side \cite{Xing09}. A straightforward way out is to
lower the seesaw scale down to the TeV scale, an energy frontier
to be soon explored by the Large Hadron Collider (LHC). But such a
TeV-scale seesaw scenario inevitably suffers from a terrible
fine-tuning of cancellations between the Yukawa coupling texture
and the heavy Majorana mass matrix \cite{KS}. To resolve this
unnaturalness problem built in the canonical (type-I) seesaw
mechanism at low energies, some new interest has recently been
paid to a relatively old idea --- the inverse seesaw mechanism
\cite{MV}.

The inverse seesaw mechanism, which can be regarded as the
simplest multiple seesaw picture \cite{Zhou09}, is an extension of
the canonical seesaw mechanism by introducing three additional
gauge-singlet neutrinos together with one gauge-singlet scalar.
Its typical result for the effective mass matrix of three active
neutrinos is $M^{}_\nu = M^{}_{\rm D}(M^T_{\rm R})^{-1} M^{}_\mu
(M^{}_{\rm R})^{-1} M^T_{\rm D}$ in the leading-order
approximation, where the scales of three mass matrices may
naturally satisfy $M^{}_{\rm R} \gg M^{}_{\rm D} \gg M^{}_\mu$.
The smallness of $M^{}_\nu$ can be attributed to both the
smallness of $M^{}_\mu$ and the smallness of $M^{}_{\rm
D}/M^{}_{\rm R}$ at the TeV scale (i.e., $M^{}_{\rm R} \sim {\cal
O}(1)$ TeV). It is therefore possible to get a balance between
theoretical naturalness and experimental testability of the
inverse seesaw scheme. Nevertheless, the inverse seesaw mechanism
itself is impossible to interpret the observed pattern of neutrino
mixing, which is composed of two large angles ($\theta^{}_{12}
\sim 34^\circ$ and $\theta^{}_{23} \sim 45^\circ$) and one small
angle ($\theta^{}_{13} < 10^\circ$) \cite{Fogli}, because the
flavor structures of $M^{}_{\rm R}$, $M^{}_{\rm D}$ and $M^{}_\mu$
are entirely unspecified. The latter can be determined by imposing
certain flavor symmetries, but such flavor symmetries usually need
to be broken in order to give rise to the correct neutrino mass
spectrum and neutrino mixing pattern. For example, a proper
combination of the $A^{}_4$ flavor symmetry and the inverse seesaw
mechanism at the TeV scale \cite{Valle} may successfully predict
the tri-bimaximal neutrino mixing pattern \cite{TBM} (with
$\theta^{}_{12} = \arctan(1/\sqrt{2}) \approx 35.3^\circ$,
$\theta^{}_{13} = 0^\circ$ and $\theta^{}_{23} = 45^\circ$), which
is definitely consistent with current experimental data on solar,
atmospheric, reactor and accelerator neutrino oscillations.

The present work aims to combine the inverse seesaw mechanism with
a newly-proposed flavor symmetry --- the Friedberg-Lee (FL)
symmetry \cite{FL}, so as to fix the flavor textures of $M^{}_{\rm
R}$, $M^{}_{\rm D}$ and $M^{}_\mu$ and thus predict the mass
spectrum and mixing pattern of three active neutrinos at the TeV
scale. As the FL symmetry requires a fermion mass term to be
invariant under a space-time-independent translation of the
relevant fermion fields, it can more reasonably be applied to the
gauge-singlet neutrino sector instead of the active neutrino
sector. We show that the FL symmetry forces one pair of the
gauge-singlet neutrinos to be massless, leading to a simplified
but viable version of the inverse seesaw mechanism which has two
pairs of massive gauge-singlet neutrinos and allows one active
neutrino to be massless. With the help of $\mu$-$\tau$ permutation
symmetry, we consider a very simple FL symmetry breaking ansatz in
the gauge-singlet neutrino sector and obtain the tri-bimaximal
neutrino mixing pattern by means of the inverse seesaw relation.
We find that non-unitary corrections to this special mixing
pattern are possible to reach the percent level if $M^{}_{\rm
D}/M^{}_{\rm R} \sim {\cal O}(0.1)$ holds at the TeV scale. We
also demonstrate that a soft breaking of $\mu$-$\tau$ symmetry can
easily accommodate CP violation in such an inverse seesaw
scenario.

\section{The inverse seesaw mechanism with FL symmetry}

Let us work in the basis where the flavor eigenstates of three
charged leptons are identified with their mass eigenstates
throughout this paper. Different from the canonical seesaw
mechanism with three right-handed neutrinos $N^{i}_{\rm R}$ (for
$i=1,2,3$), the inverse seesaw scheme contains three additional
gauge-singlet neutrinos $S^{i}_{\rm R}$ (for $i=1,2,3$) together
with one gauge-singlet scalar $\Phi$. Allowing for lepton number
violation to a certain extent, one can write out the following
gauge-invariant neutrino mass terms in the inverse seesaw
mechanism:
\begin{eqnarray}
-{\cal L}^{}_\nu = \overline{\ell^{}_{\rm L}} Y^{}_\nu \tilde{H}
N^{}_{\rm R} + \overline{N^c_{\rm R}} Y^\prime_\nu S^{}_{\rm R}
\Phi + \frac{1}{2} \overline{S^c_{\rm R}} M^{}_\mu S^{}_{\rm R} +
{\rm h.c.} \; ,
\end{eqnarray}
where $\ell^{}_{\rm L}$ and $\tilde{H} \equiv i\sigma^{}_2 H^*$
stand respectively for the $SU(2)^{}_{\rm L}$ lepton and Higgs
doublets, $Y^{}_\nu$ and $Y^\prime_\nu$ are the $3\times 3$ Yukawa
coupling matrices, and $M^{}_\mu$ is a symmetric Majorana mass
matrix. After spontaneous gauge symmetry breaking, Eq. (1) becomes
\begin{eqnarray}
-{\cal L}^\prime_\nu = \overline{\nu^{}_{\rm L}} M^{}_{\rm D}
N^{}_{\rm R} + \overline{N^c_{\rm R}} M^{}_{\rm R} S^{}_{\rm R} +
\frac{1}{2} \overline{S^c_{\rm R}} M^{}_\mu S^{}_{\rm R} + {\rm
h.c.} \; ,
\end{eqnarray}
where $M^{}_{\rm D} \equiv Y^{}_\nu \langle H \rangle$ and
$M^{}_{\rm R} = Y^\prime_\nu \langle \Phi \rangle$ are the
$3\times 3$ mass matrices. Then we arrive at the overall $9 \times
9$ neutrino mass matrix $\cal M$ in the flavor basis defined by
$(\nu^{}_{\rm L}, N^c_{\rm R}, S^c_{\rm R})$ and their
charge-conjugate states:
\begin{eqnarray}
{\cal M} = \left(\matrix{{\bf 0} & M^{}_{\rm D} & {\bf 0} \cr
M^T_{\rm D} & {\bf 0} & M^{}_{\rm R} \cr {\bf 0} & M^T_{\rm R} &
M^{}_\mu \cr} \right) \; .
\end{eqnarray}
Note that the interesting nearest-neighbor-interaction pattern of
$\cal M$ is guaranteed by an implementation of the global
$U(1)\times Z^{}_4$ symmetry with a proper charge assignment
\cite{Zhou09}. Note also that the mass scales of three
sub-matrices in $\cal M$ may naturally have a hierarchy $M^{}_{\rm
R} \gg M^{}_{\rm D} \gg M^{}_\mu$, because the second mass term in
${\cal L}^\prime_\nu$ is not subject to the $SU(2)^{}_{\rm L}$
gauge symmetry breaking scale and the third mass term in ${\cal
L}^\prime_\nu$ violates the lepton number conservation
\cite{Hooft}. In the leading-order approximation, one obtains the
inverse seesaw relation for the effective mass matrix of three
active neutrinos:
\begin{eqnarray}
M^{}_\nu = M^{}_{\rm D} (M^T_{\rm R})^{-1} M^{}_\mu (M^{}_{\rm
R})^{-1} M^T_{\rm D} \; .
\end{eqnarray}
It becomes obvious that $M^{}_\nu \rightarrow {\bf 0}$ holds in
the limit $M^{}_\mu \rightarrow {\bf 0}$. For instance, $M^{}_\nu
\sim {\cal O}(0.1)$ eV can easily be achieved from $M^{}_{\rm
D}/M^{}_{\rm R} \sim {\cal O}(10^{-2})$ and $M^{}_\mu \sim {\cal
O} (1)$ keV.

Without loss of generality, let us work in a basis where
$Y^\prime_\nu$ (or equivalently, $M^{}_{\rm R}$) is diagonal;
i.e., $Y^\prime_\nu = {\rm Diag}\{y^\prime_1, y^\prime_2,
y^\prime_3\}$ with $y^\prime_i \equiv M^{}_i/\langle \Phi\rangle$
and $M^{}_i$ (for $i=1,2,3$) being real and positive. Now we
impose the FL translation \cite{FL} on both $N^{}_{\rm R}$ and
$S^{}_{\rm R}$ fields:
\begin{eqnarray}
N^{i}_{\rm R} \rightarrow N^{i}_{\rm R} + \xi^{}_i \Theta \; ,
~~~~ S^{i}_{\rm R} \rightarrow S^{i}_{\rm R} + \xi^{}_i \Theta \;
,
\end{eqnarray}
where $\xi^{}_i$ (for $i=1,2,3$) are in general complex numbers,
and $\Theta$ is a space-time-independent Grassmann parameter
(i.e., $\Theta$ is anti-commuting and thus $\Theta^2 =0$ holds).
Note that the gauge-singlet neutrinos $N^{i}_{\rm R}$ and
$S^{i}_{\rm R}$ do not interact with the gauge bosons of the
standard model. Note also that the kinetic terms of $N^{i}_{\rm
R}$ and $S^{i}_{\rm R}$ change under the above translation, but
the resulting action is invariant just because $\Theta$ is
independent of space and time \cite{J08}. Hence the whole
Lagrangian of an inverse seesaw model will be invariant under the
FL translation, if and only if we require three neutrino mass
terms in ${\cal L}^{}_\nu$ to be invariant under the FL
translation. This requirement simply implies
\begin{eqnarray}
(Y^{}_\nu)^{}_{ij} \xi^{}_j = 0 \; ; ~~~~ \xi^{}_i
(Y^\prime_\nu)^{}_{ij} = 0 \; , ~~~~ (Y^\prime_\nu)^{}_{ij}
\xi^{}_j = 0 \; ; ~~~~ \xi^{}_i (M^{}_\mu)^{}_{ij} = 0 \; , ~~~~
(M^{}_\mu)^{}_{ij} \xi^{}_j = 0
\end{eqnarray}
for $i,j = 1, 2, 3$. In other words, each mass matrix must have a
zero eigenvalue. Because $Y^\prime_\nu$ is diagonal, we may simply
choose $y^\prime_3 =0$ or equivalently
\begin{eqnarray}
Y^\prime_\nu = \left( \matrix{y^\prime_1 & 0 & 0 \cr 0 &
y^\prime_2 & 0 \cr 0 & 0 & 0 \cr} \right) \; ,
\end{eqnarray}
such that $\xi^{}_1 = \xi^{}_2 = 0$ and $\xi^{}_3 \neq 0$ must
hold to satisfy the second and third conditions given in Eq. (6).
This in turn implies that the elements in the third column of
$Y^{}_\nu$ and those in the third row and the third column of
$M^{}_\mu$ must be vanishing, so as to satisfy the first, fourth
and fifth conditions shown in Eq. (6). The textures of $Y^{}_\nu$
and $M^{}_\mu$ are therefore given as
\begin{eqnarray}
Y^{}_\nu & = & \left( \matrix{ y^{}_{11} & ~y^{}_{12} & 0 \cr
y^{}_{21} & ~y^{}_{22} & 0 \cr y^{}_{31} & ~y^{}_{32} & 0 \cr}
\right) \; , \nonumber \\
M^{}_\mu & = & \left( \matrix{ \mu^{}_{11} & \mu^{}_{12} & 0 \cr
\mu^{}_{12} & \mu^{}_{22} & 0 \cr 0 & 0 & 0 \cr} \right) \; .
\end{eqnarray}
In this case, the symmetric $9\times 9$ neutrino mass matrix $\cal
M$ in Eq. (3) can be simplified to an effective $7\times 7$
neutrino mass matrix
\begin{eqnarray}
{\cal M} = \left( \matrix{ 0 & 0 & 0 & \times & \times & 0 & 0 \cr
0 & 0 & 0 & \times & \times & 0 & 0 \cr 0 & 0 & 0 & \times &
\times & 0 & 0 \cr \times & \times & \times & 0 & 0 & \times & 0
\cr \times & \times & \times & 0 & 0 & 0 & \times \cr 0 & 0 & 0 &
\times & 0 & \times & \times \cr 0 & 0 & 0 & 0 & \times & \times &
\times \cr} \right) \; ,
\end{eqnarray}
where ``$\times$" symbolically denotes an arbitrary (non-zero)
matrix element. Because ${\rm Det}{\cal M} = 0$ holds, one active
neutrino must be massless in this minimal inverse seesaw scenario.
The phenomenology of such a simple scenario has partly been
explored in Ref. \cite{Zhang}. Here we show that the FL symmetry
applied to the gauge-singlet neutrino sector naturally explains
why a pair of gauge-singlet neutrinos can be decoupled from the
usual inverse seesaw mechanism.

It is worth remarking that the above conclusion does not depend on
the chosen basis of $Y^\prime_\nu$ (i.e., $Y^\prime_\nu$ is
diagonal), since a unitary transformation of either $N^{}_{\rm R}$
or $S^{}_{\rm R}$ does not change any physical content of the
inverse seesaw mechanism. To derive a phenomenologically-favored
neutrino mixing pattern from this interesting scheme, however, it
is more convenient to introduce a less stringent FL symmetry into
the gauge-singlet neutrino sector. Here we consider the case of
$\xi^{}_1 = \xi^{}_2 = \xi^{}_3 \equiv \xi$, which is equivalent
to the original FL translation (imposed on $\nu^{}_{\rm L}$
\cite{FL,XZZ,LX,LXL}). Then three mass terms in ${\cal
L}^\prime_\nu$ can keep invariant under the translations
$N^{i}_{\rm R} \rightarrow N^{i}_{\rm R} + \xi \Theta$ and
$S^{i}_{\rm R} \rightarrow S^{i}_{\rm R} + \xi \Theta$ if they
take the following forms:
\begin{eqnarray}
&& \sum_{\alpha, i} \overline{\nu^{}_{\alpha\rm L}} (M^{}_{\rm
D})^{}_{\alpha i} N^{i}_{\rm R} = \sum_\alpha A^{}_\alpha
\overline{\nu^{}_{\alpha\rm L}} (N^{3}_{\rm R} - N^{2}_{\rm R}) +
\sum_\alpha B^{}_\alpha \overline{\nu^{}_{\alpha\rm L}}
(N^{2}_{\rm R} - N^{1}_{\rm R}) + \sum_\alpha C^{}_\alpha
\overline{\nu^{}_{\alpha\rm L}} (N^{1}_{\rm R} - N^{3}_{\rm R}) \;
, \nonumber \\ \nonumber \\
&& \sum_{i, j} \overline{N^{i c}_{\rm R}}
(M^{}_{\rm R})^{}_{ij} S^{j}_{\rm R} = A (\overline{N^{3 c}_{\rm
R}} - \overline{N^{2 c}_{\rm R}}) (S^{3}_{\rm R} - S^{2}_{\rm R})
+ B (\overline{N^{2 c}_{\rm R}} - \overline{N^{1 c}_{\rm R}})
(S^{2}_{\rm R} -
S^{1}_{\rm R}) \nonumber \\
&& ~~~~~~~~~~~~~~~~~~~~~~~~~ + C (\overline{N^{1 c}_{\rm R}} -
\overline{N^{3 c}_{\rm R}}) (S^{1}_{\rm R} - S^{3}_{\rm R}) \; ,
\nonumber \\ \nonumber \\
&& \sum_{i, j} \overline{S^{i c}_{\rm R}} (M^{}_\mu)^{}_{ij}
S^{j}_{\rm R} = a (\overline{S^{3 c}_{\rm R}} - \overline{S^{2
c}_{\rm R}}) (S^{3}_{\rm R} - S^{2}_{\rm R}) + b (\overline{S^{2
c}_{\rm R}} - \overline{S^{1 c}_{\rm R}}) (S^{2}_{\rm R} -
S^{1}_{\rm R}) \nonumber \\
&& ~~~~~~~~~~~~~~~~~~~~~~~~ + c (\overline{S^{1 c}_{\rm R}} -
\overline{S^{3 c}_{\rm R}}) (S^{1}_{\rm R} - S^{3}_{\rm R}) \; ,
\end{eqnarray}
where the Greek index runs over $(e, \mu, \tau)$ and the Latin
index runs over $(1, 2, 3)$. The explicit expressions of
$M^{}_{\rm D}$, $M^{}_{\rm R}$ and $M^{}_\mu$ turn out to be
\begin{eqnarray}
M^{}_{\rm D} & = & \left( \matrix{ C^{}_e - B^{}_e & B^{}_e -
A^{}_e & A^{}_e - C^{}_e \cr C^{}_\mu - B^{}_\mu & B^{}_\mu -
A^{}_\mu & A^{}_\mu - C^{}_\mu \cr C^{}_\tau - B^{}_\tau &
B^{}_\tau - A^{}_\tau & A^{}_\tau - C^{}_\tau \cr} \right) \; ,
\nonumber \\
M^{}_{\rm R} & = & \left( \matrix{ B + C & -B & -C \cr -B & A+B &
-A \cr -C & -A & A+C \cr}\right) \; , \nonumber \\
M^{}_\mu & = & \left( \matrix{ b + c & -b & -c \cr -b & a+b & -a
\cr -c & -a & a+c \cr}\right) \; ,
\end{eqnarray}
in which all the matrix elements are in general complex. It is
easy to check that ${\rm Det}M^{}_{\rm D} = {\rm Det}M^{}_{\rm R}
= {\rm Det}M^{}_\mu = 0$ holds, and thus each mass matrix has one
zero eigenvalue. We see that the above FL symmetry must be broken,
at least for the second mass term of ${\cal L}^\prime$, such that
${\rm Det}M^{}_{\rm R} \neq 0$ holds to make the inverse seesaw
formula in Eq. (4) applicable.

\section{A FL symmetry breaking ansatz with $\mu$-$\tau$ symmetry}

There are certainly a variety of possibilities of breaking the FL
symmetry. Here we follow the spirit of Ref. \cite{FL} and consider
a very simple symmetry breaking ansatz for the second and third
mass terms in Eq. (10):
\begin{eqnarray}
&& \sum_{i, j} \overline{N^{i c}_{\rm R}} (M^{}_{\rm R})^{}_{ij}
S^{j}_{\rm R} \rightarrow \sum_{i, j} \overline{N^{i c}_{\rm R}}
(M^{}_{\rm R})^{}_{ij} S^{j}_{\rm R} + M^{}_0 \sum_i
\overline{N^{i c}_{\rm R}} S^{i}_{\rm R} \; , \nonumber \\
&& \sum_{i, j} \overline{S^{i c}_{\rm R}} (M^{}_\mu)^{}_{ij}
S^{j}_{\rm R} \rightarrow \sum_{i, j} \overline{S^{i c}_{\rm R}}
(M^{}_\mu)^{}_{ij} S^{j}_{\rm R} + \mu^{}_0 \sum_i \overline{S^{i
c}_{\rm R}} S^{i}_{\rm R} \; ,
\end{eqnarray}
where $M^{}_0$ and $\mu^{}_0$ are real and positive. To minimize
the number of free parameters, we impose the $\mu$-$\tau$
permutation symmetry on these two mass terms (i.e., $B=C$ and
$b=c$) and assume all of their parameters to be real. The
resultant mass matrices read
\begin{eqnarray}
M^{}_{\rm R} & = & M^{}_0 \left[ \left( \matrix{ 1 & 0 & 0 \cr 0 &
1 & 0 \cr 0 & 0 & 1 \cr} \right) + \left( \matrix{ 2B^\prime &
-B^\prime & -B^\prime \cr -B^\prime & A^\prime + B^\prime &
-A^\prime \cr -B^\prime & -A^\prime & A^\prime + B^\prime \cr}
\right) \right] \; ,
\nonumber \\
M^{}_\mu & = & \mu^{}_0 \left[ \left( \matrix{ 1 & 0 & 0 \cr 0 & 1
& 0 \cr 0 & 0 & 1 \cr} \right) + \left( \matrix{ 2b^\prime &
-b^\prime & -b^\prime \cr -b^\prime & a^\prime + b^\prime &
-a^\prime \cr -b^\prime & -a^\prime & a^\prime + b^\prime \cr}
\right) \right] \; ,
\end{eqnarray}
in which $A^\prime \equiv A/M^{}_0$ (or $a^\prime \equiv
a/\mu^{}_0$) and $B^\prime \equiv B/M^{}_0$ (or $b^\prime \equiv
b/\mu^{}_0$) are defined to be two dimensionless parameters. The
diagonalization of $M^{}_{\rm R}$ or $M^{}_\mu$ is rather
straightforward: $V^T_0 M^{}_{\rm R} V^{}_0 = {\rm Diag}\{M^{}_1,
M^{}_2, M^{}_3\}$ and $V^T_0 M^{}_\mu V^{}_0 = {\rm
Diag}\{\mu^{}_1, \mu^{}_2, \mu^{}_3\}$, where $V^{}_0$ is simply
the tri-bimaximal mixing pattern \cite{TBM}
\begin{eqnarray}
V^{}_0 = \left( \matrix{ \displaystyle\frac{2}{\sqrt{6}} &
\displaystyle\frac{1}{\sqrt{3}} & 0 \cr \displaystyle
-\frac{1}{\sqrt{6}} & \displaystyle \frac{1}{\sqrt{3}} &
\displaystyle \frac{1}{\sqrt{2}} \cr \displaystyle
-\frac{1}{\sqrt{6}} & \displaystyle \frac{1}{\sqrt{3}} &
\displaystyle -\frac{1}{\sqrt{2}} \cr} \right) \; .
\end{eqnarray}
Then we obtain three mass eigenvalues $M^{}_1 = M^{}_0 (1 +
3B^\prime)$, $M^{}_2 = M^{}_0$, $M^{}_3 = M^{}_0 (1 + 2A^\prime +
B^\prime)$; and similarly $\mu^{}_1 = \mu^{}_0 (1 + 3b^\prime)$,
$\mu^{}_2 = \mu^{}_0$, $\mu^{}_3 = \mu^{}_0 (1 + 2a^\prime +
b^\prime)$. In terms of mass eigenvalues, $M^{}_{\rm R}$ and
$M^{}_\mu$ can be reexpressed as
\begin{eqnarray}
M^{}_{\rm R} = M^{}_1 X + M^{}_2 Y + M^{}_3 Z \; , ~~~~ M^{}_\mu =
\mu^{}_1 X + \mu^{}_2 Y + \mu^{}_3 Z \; ,
\end{eqnarray}
where
\begin{eqnarray}
X & = & \frac{1}{6} \left( \matrix{ 4 & -2 & -2 \cr -2 & 1 & 1 \cr
-2 & 1 & 1 \cr} \right) \; , \nonumber \\
Y & = & \frac{1}{3} \left( \matrix{ ~1 & ~~~1~~ & 1~ \cr ~1 &
~~~1~~ & 1~ \cr ~1 & ~~~1~~ & 1~ \cr} \right) \; , \nonumber \\
Z & = & \frac{1}{2} \left( \matrix{ ~0~ & ~0 & 0 \cr ~0~ & ~1 & -1
\cr ~0~ & -1& 1 \cr} \right) \; .
\end{eqnarray}
Note that $X^2 = X$, $Y^2 = Y$, $Z^2 = Z$ and $XY = YX = XZ = ZX =
YZ = ZY = {\bf 0}$ hold. Note also that the inverse matrix of
$M^{}_{\rm R}$ takes the same form as $M^{}_{\rm R}$ itself:
\begin{eqnarray}
(M^{}_{\rm R})^{-1} = M^{-1}_1 X + M^{-1}_2 Y + M^{-1}_3 Z \; .
\end{eqnarray}
Now let us make a purely phenomenological assumption: $M^{}_{\rm
D} = v {\bf 1}$ with $v \sim \langle H\rangle \approx 174$ GeV
being the electroweak scale and $\bf 1$ being the identity matrix,
just for the sake of simplicity. Then the inverse seesaw formula
in Eq. (4) allows us to arrive at
\begin{eqnarray}
M^{}_\nu = \frac{v^2 \mu^{}_1}{M^2_1} X + \frac{v^2
\mu^{}_2}{M^2_2} Y + \frac{v^2 \mu^{}_3}{M^2_3} Z \; .
\end{eqnarray}
It is straightforward to show that this effective mass matrix can
also be diagonalized by the unitary transformation $V^T_0 M^{}_\nu
V^{}_0 = {\rm Diag}\{m^{}_1, m^{}_2, m^{}_3\}$, where $V^{}_0$ has
been given in Eq. (14) and three neutrino masses $m^{}_i = v^2
\mu^{}_i/M^2_i$ (for $i=1,2,3$) directly reflect the salient
feature of the inverse seesaw mechanism. In other words, the
smallness of $m^{}_i$ is ascribed to both the smallness of
$\mu^{}_i$ and that of $v^2/M^2_i$. Current neutrino oscillation
data only provide us with $\Delta m^2_{21} \equiv m^2_2 - m^2_1
\approx 7.7 \times 10^{-5} ~ {\rm eV}^2$ and $m^2_{32} \equiv
m^2_3 - m^2_2 \approx \pm 2.4 \times 10^{-3} ~ {\rm eV}^2$
\cite{Fogli}, and thus these two neutrino mass-squared differences
can easily be reproduced from our result for $m^{}_i$ by adjusting
six free parameters $\mu^{}_i$ and $M^{}_i$ (for $i=1,2,3$).

Note that the neutrino mixing matrix $V$ appearing in the
charged-current interactions of three active neutrinos is not
exactly $V^{}_0$ used to diagonalize $M^{}_\nu$ in Eq. (18), just
because of slight mixing between light and heavy neutrinos in the
inverse seesaw mechanism. As shown in Appendix A, the
charged-current interactions of three light Majorana neutrinos
read
\begin{eqnarray}
-{\cal L}^{}_{\rm cc} = \frac{g}{\sqrt{2}}
\overline{\left(\matrix{e & \mu & \tau}\right)^{}_{\rm L}}
~\gamma^\mu ~V \left( \matrix{\nu^{}_1 \cr \nu^{}_2 \cr \nu^{}_3}
\right)^{}_{\rm L} W^-_\mu + {\rm h.c.} \; ,
\end{eqnarray}
in which $(e, \mu, \tau)$ and $(\nu^{}_1, \nu^{}_2, \nu^{}_3)$ are
the mass eigenstates of charged leptons and active neutrinos,
respectively. The relationship between $V$ and $V^{}_0$ is given
by $V = ({\bf 1} -\eta) V^{}_0$, where $\eta \lesssim {\cal
O}(10^{-2})$ signifies the slight deviation of $V$ from $V^{}_0$
and its approximate expression can be found in Eq. (A7). For the
ansatz under consideration, we explicitly obtain
\begin{eqnarray}
\eta \approx \frac{1}{2} M^{}_{\rm D} (M^{}_{\rm R})^{-2}
M^{}_{\rm D} = \frac{1}{2} \left[ \frac{v^2}{M^2_1} X +
\frac{v^2}{M^2_2} Y + \frac{v^2}{M^2_3} Z \right] \; .
\end{eqnarray}
Current experimental constraints on the matrix elements of $\eta$
are
\begin{eqnarray}
\left|\eta\right| < \left(\matrix{ 5.5 \times 10^{-3} & 3.5 \times
10^{-5} & 8.0 \times 10^{-3} \cr 3.5 \times 10^{-5} & 5.0 \times
10^{-3} & 5.0 \times 10^{-3} \cr 8.0 \times 10^{-3} & 5.0 \times
10^{-3} & 5.0 \times 10^{-3} \cr}\right) \; ,
\end{eqnarray}
at the $90\%$ confidence level \cite{Antusch}. Combining Eqs.
(16), (20) and (21), we arrive at
\begin{eqnarray}
\left| \eta^{}_{ee} \right| & \approx & \frac{1}{6} \left|
\frac{2v^2}{M^2_1} + \frac{v^2}{M^2_2} \right| < 5.5 \times
10^{-3} \; , \nonumber \\
\left| \eta^{}_{e\mu} \right| & \approx & \frac{1}{6} \left|
\frac{v^2}{M^2_1} - \frac{v^2}{M^2_2} \right| < 3.5 \times 10^{-5}
\; , \nonumber \\
\left| \eta^{}_{\mu\mu} \right| & \approx & \frac{1}{12} \left|
\frac{v^2}{M^2_1} + \frac{2v^2}{M^2_2} + \frac{3v^2}{M^2_3}
\right| < 5.0 \times 10^{-3} \; , \nonumber \\
\left| \eta^{}_{\mu\tau} \right| & \approx & \frac{1}{12} \left|
\frac{v^2}{M^2_1} + \frac{2v^2}{M^2_2} - \frac{3v^2}{M^2_3}
\right| < 5.0 \times 10^{-3} \; ,
\end{eqnarray}
together with $|\eta^{}_{e\tau}| = |\eta^{}_{e\mu}|$ and
$|\eta^{}_{\tau\tau}| = |\eta^{}_{\mu\mu}|$ due to the
$\mu$-$\tau$ symmetry of Hermitian $\eta$. If $M^{}_i \sim {\cal
O}(1)$ TeV, then $M^{}_1 \approx M^{}_2$ is expected from the
stringent constraint on $|\eta^{}_{e\mu}|$. This result
accordingly implies $|\eta^{}_{ee}| \approx v^2/M^2_1 < 1.1 \times
10^{-2}$, from which $M^{}_1 > 1.6$ TeV can be extracted. It is in
general difficult to get a limit on $M^{}_3$. But if $M^{}_1
\approx M^{}_2 \approx 1.8$ TeV is taken, for example, then
$|\eta^{}_{\mu\mu}| < 5.0 \times 10^{-3}$ leads us to $M^{}_3
> 1.7$ TeV.

With the help of Eqs. (14) and (20), it is straightforward to
obtain an interesting expression of $V = ({\bf 1} -\eta)V^{}_0$ in
this inverse seesaw scenario:
\begin{eqnarray}
V = \left( \matrix{ \displaystyle\frac{2}{\sqrt{6}} &
\displaystyle\frac{1}{\sqrt{3}} & 0 \cr \displaystyle
-\frac{1}{\sqrt{6}} & \displaystyle \frac{1}{\sqrt{3}} &
\displaystyle \frac{1}{\sqrt{2}} \cr \displaystyle
-\frac{1}{\sqrt{6}} & \displaystyle \frac{1}{\sqrt{3}} &
\displaystyle -\frac{1}{\sqrt{2}} \cr} \right) \left( \matrix{
\displaystyle 1 - \frac{v^2}{2M^2_1} & 0 & 0 \cr 0 & \displaystyle
1 - \frac{v^2}{2M^2_2} & 0 \cr 0 & 0 & \displaystyle 1 -
\frac{v^2}{2M^2_3} \cr} \right) \; .
\end{eqnarray}
This simple but instructive result clearly shows the deviation of
$V$ from $V^{}_0$. In particular, $V^{}_{e3} = (V^{}_0)^{}_{e3} =
0$ holds; i.e., the non-unitary effect does not contribute to this
smallest neutrino mixing matrix element.

We remark that the above FL symmetry breaking ansatz is viable and
suggestive for building a realistic inverse seesaw model at the
TeV scale. To generate non-vanishing $V^{}_{e3}$ and CP violation,
however, one has to invoke a different FL symmetry breaking
pattern with one or more non-trivial $CP$-violating phases.

\section{Soft $\mu$-$\tau$ symmetry breaking and CP violation}

We proceed with the FL symmetry breaking ansatz in Eq. (12) but
allow soft $\mu$-$\tau$ symmetry breaking for $M^{}_{\rm R}$ and
$M^{}_\mu$ (i.e., $C = B^*$ and $c = b^*$ with $B$ and $b$ being
complex):
\begin{eqnarray}
M^{}_{\rm R} & = & M^{}_0 \left[ \left( \matrix{ 1 & 0 & 0 \cr 0 &
1 & 0 \cr 0 & 0 & 1 \cr} \right) + \left( \matrix{ 2{\rm
Re}B^\prime & -B^\prime & -{B^\prime}^* \cr -B^\prime & A^\prime +
B^\prime & -A^\prime \cr -{B^\prime}^* & -A^\prime & A^\prime +
{B^\prime}^* \cr} \right) \right] \; ,
\nonumber \\
M^{}_\mu & = & \mu^{}_0 \left[ \left( \matrix{ 1 & 0 & 0 \cr 0 & 1
& 0 \cr 0 & 0 & 1 \cr} \right) + \left( \matrix{ 2{\rm Re}b^\prime
& -b^\prime & -{b^\prime}^* \cr -b^\prime & a^\prime + b^\prime &
-a^\prime \cr -{b^\prime}^* & -a^\prime & a^\prime + {b^\prime}^*
\cr} \right) \right] \; ,
\end{eqnarray}
where $M^{}_0$ (or $\mu^{}_0$) and $A^\prime$ (or $a^\prime$) are
real. As shown in Appendix B, the inverse matrix of $M^{}_{\rm R}$
takes the same texture as $M^{}_{\rm R}$ itself:
\begin{eqnarray}
(M^{}_{\rm R})^{-1} = \frac{1}{M^{}_0} \left[ \left( \matrix{ 1 &
0 & 0 \cr 0 & 1 & 0 \cr 0 & 0 & 1 \cr} \right) + \left( \matrix{
2{\rm Re}B^{\prime\prime} & -B^{\prime\prime} &
-{B^{\prime\prime}}^* \cr -B^{\prime\prime} & A^{\prime\prime} +
B^{\prime\prime} & -A^{\prime\prime} \cr -{B^{\prime\prime}}^* &
-A^{\prime\prime} & A^{\prime\prime} + {B^{\prime\prime}}^* \cr}
\right) \right] \; ,
\end{eqnarray}
where
\begin{eqnarray}
A^{\prime\prime} & = & - \frac{A^\prime + (2A^\prime {\rm Re}
B^\prime + |B^\prime|^2)}{1 + 2 (A^\prime + 2{\rm Re} B^\prime) +
3 (2 A^\prime {\rm Re}B^\prime + |B^\prime|^2)} \;\; , \nonumber \\
B^{\prime\prime} & = & - \frac{B^\prime + (2A^\prime {\rm Re}
B^\prime + |B^\prime|^2)}{1 + 2 (A^\prime + 2{\rm Re} B^\prime) +
3 (2 A^\prime {\rm Re}B^\prime + |B^\prime|^2)} \;\; ,
\end{eqnarray}
which can easily be read off from Eq. (B3) in Appendix B.
Furthermore, we show that $(M^T_{\rm R})^{-1} M^{}_\mu (M^{}_{\rm
R})^{-1}$ is also of the same texture:
\begin{eqnarray}
(M^T_{\rm R})^{-1} M^{}_\mu (M^{}_{\rm R})^{-1} & = &
\frac{\mu^{}_0}{M^2_0} \left[ \left( \matrix{ 1 & 0 & 0 \cr 0 & 1
& 0 \cr 0 & 0 & 1 \cr} \right) + \left( \matrix{ 2{\rm Re}
\widehat{B} & -\widehat{B} & -\widehat{B}^* \cr -\widehat{B} &
\widehat{A} + \widehat{B} & -\widehat{A} \cr -\widehat{B}^* &
-\widehat{A} & \widehat{A} + \widehat{B}^* \cr} \right) \right] \;
,
\end{eqnarray}
where
\begin{eqnarray}
\widehat{A} & = & + A^{\prime\prime} \left[ \left( 1 +
A^{\prime\prime} + B^{\prime\prime} \right) \left( 1 + a^\prime +
b^\prime \right) + A^{\prime\prime} a^\prime + B^{\prime\prime}
b^\prime \right] \nonumber \\
&& - {B^{\prime\prime}}^* \left[
\left( 1 + A^{\prime\prime} + B^{\prime\prime} \right) b^\prime +
B^{\prime\prime} \left( 1 + 2{\rm Re} b^\prime \right) -
A^{\prime\prime} {b^\prime}^* \right] \nonumber \\
&& + \left( 1 + A^{\prime\prime} + {B^{\prime\prime}}^* \right)
\left[ \left( 1 + A^{\prime\prime} + B^{\prime\prime} \right)
a^\prime + A^{\prime\prime} \left( 1 + a^\prime + {b^\prime}^*
\right) - B^{\prime\prime} {b^\prime}^* \right] \; , \nonumber \\
\widehat{B} & = & -A^{\prime\prime} \left[ \left( 1 + 2{\rm Re}
B^{\prime\prime}\right) {b^\prime}^* + {B^{\prime\prime}}^* \left(
1 + a^\prime + {b^\prime}^* \right) - B^{\prime\prime} a^\prime
\right] \nonumber \\
&& + B^{\prime\prime} \left[ \left(1 + 2{\rm
Re}B^{\prime\prime}\right) \left(1 + 2{\rm Re}b^\prime \right) +
B^{\prime\prime} b^\prime + {B^{\prime\prime}}^* {b^\prime}^* \right] \nonumber \\
&& + \left( 1 + A^{\prime\prime} + B^{\prime\prime} \right) \left[
\left( 1 + 2{\rm Re}B^{\prime\prime} \right) b^\prime +
B^{\prime\prime} \left( 1 + a^\prime + b^\prime \right) -
{B^{\prime\prime}}^* a^\prime \right] \; ,
\end{eqnarray}
which can directly be read off from Eq. (B6). Note that $\widehat
A$ is real and $\widehat B$ is complex. Making the same
phenomenological assumption for $M^{}_{\rm D}$ as in section III
(i.e., $M^{}_{\rm D} = v {\bf 1}$), we simply obtain the effective
mass matrix of three active neutrinos from Eqs. (4) and (27) in
this inverse seesaw scenario:
\begin{eqnarray}
M^{}_\nu = m^{}_0 \left[ \left( \matrix{ 1 & 0 & 0 \cr 0 & 1 & 0
\cr 0 & 0 & 1 \cr} \right) + \left( \matrix{ 2{\rm Re} \widehat{B}
& -\widehat{B} & -\widehat{B}^* \cr -\widehat{B} & \widehat{A} +
\widehat{B} & -\widehat{A} \cr -\widehat{B}^* & -\widehat{A} &
\widehat{A} + \widehat{B}^* \cr} \right) \right]
\end{eqnarray}
with $m^{}_0 = v^2\mu^{}_0/M^2_0$. The small mass eigenvalues of
$M^{}_\nu$ are therefore attributed to small $\mu^{}_0$ as well as
small $v^2/M^2_0$ at the scale of $M^{}_0 \sim {\cal O}(1)$ TeV.

The diagonalization of $M^{}_\nu$ in Eq. (29) can be done by using
the unitary transformation $U^\dagger_0 M^{}_\nu U^*_0 = {\rm
Diag} \{m^{}_1, m^{}_2, m^{}_3 \}$, where
\begin{eqnarray}
U^{}_0 = \left( \matrix{ \displaystyle\frac{2}{\sqrt{6}} &
\displaystyle\frac{1}{\sqrt{3}} & 0 \cr \displaystyle
-\frac{1}{\sqrt{6}} & \displaystyle \frac{1}{\sqrt{3}} &
\displaystyle \frac{1}{\sqrt{2}} \cr \displaystyle
-\frac{1}{\sqrt{6}} & \displaystyle \frac{1}{\sqrt{3}} &
\displaystyle -\frac{1}{\sqrt{2}} \cr} \right) \left( \matrix{
\cos\theta & 0 & -i\sin\theta \cr 0 & 1 & 0 \cr -i\sin\theta & 0 &
\cos\theta \cr} \right) \; ,
\end{eqnarray}
and $\tan 2\theta = \sqrt{3} ~{\rm Im}\widehat{B}/(1 + \widehat{A}
+ 2{\rm Re}\widehat{B})$ arises from the soft $\mu$-$\tau$
symmetry breaking term of $M^{}_\nu$ (i.e., ${\rm Im}\widehat{B}
\neq 0$). Comparing this result with the standard parametrization
of $U^{}_0$ in terms of three mixing angles $(\theta^{}_{12},
\theta^{}_{13}, \theta^{}_{23})$ and three CP-violating phases
$(\delta, \rho, \sigma)$ \cite{PDG08,FX01}, we immediately find
\begin{eqnarray}
\theta^{}_{12} & = & \arcsin\left(\frac{1}{\sqrt{2 + \cos
2\theta}} \right) \; , \nonumber \\
\theta^{}_{13} & = & \arcsin\left( \frac{2}{\sqrt{6}}\sin\theta
\right) \; ,
\end{eqnarray}
together with $\theta^{}_{23} = 45^\circ$, $\delta = 90^\circ$ and
$\rho =\sigma =0^\circ$. On the other hand, three mass eigenvalues
of $M^{}_\nu$ are given by
\begin{eqnarray}
m^{}_1 & = & m^{}_0 \left[\sqrt{\left( 1 + \widehat{A} + 2{\rm
Re}\widehat{B}\right)^2 + 3 \left({\rm Im}\widehat{B}\right)^2} ~
- \widehat{A} + {\rm Re}\widehat{B} \right] \; , \nonumber \\
m^{}_2 & = & m^{}_0 \; , \nonumber \\
m^{}_3 & = & m^{}_0 \left[\sqrt{\left( 1 + \widehat{A} + 2{\rm
Re}\widehat{B}\right)^2 + 3 \left({\rm Im}\widehat{B}\right)^2} ~
+ \widehat{A} - {\rm Re}\widehat{B} \right] \; .
\end{eqnarray}
By adjusting four real parameters in Eq. (32), we can easily fit
two observed neutrino mass-squared differences $\Delta m^2_{21}$
and $\Delta m^2_{32}$. For example, $m^{}_0 \approx 8.8 \times
10^{-3}$ eV, $\widehat{A} \approx 2.3$, ${\rm Re}\widehat{B}
\approx -0.3$ and ${\rm Im}\widehat{B} \approx 0.3$ lead to a
hierarchical neutrino mass spectrum: $m^{}_1 \approx 0.15 m^{}_0
\approx 1.3 \times 10^{-3}$ eV, $m^{}_2 = m^{}_0 \approx 8.8
\times 10^{-3}$ eV and $m^{}_3 \approx 5.4 m^{}_0 \approx 4.8
\times 10^{-2}$ eV, consistent with current neutrino oscillation
data \cite{Fogli}. In this illustrative case, we can also obtain
$\theta \approx 6^\circ$, which in turn predicts $\theta^{}_{12}
\approx 39^\circ$ and $\theta^{}_{13} \approx 5^\circ$.

It is worth reiterating that the imaginary parts of $B^\prime$,
$b^\prime$ and $\widehat{B}$ represent the $\mu$-$\tau$ symmetry
breaking effects of $M^{}_{\rm R}$, $M^{}_\mu$ and $M^{}_\nu$,
respectively, in this phenomenological ansatz. They have nothing
to do with the $\mu$-$\tau$ symmetry breaking effect in the
charged-lepton sector, because the latter is characterized by the
mass ratio $m^{}_\tau/m^{}_\mu \approx 17$ in the chosen basis
where the charge-lepton mass matrix is diagonal and real.

Now we calculate the non-unitary corrections to $U^{}_0$ in order
to figure out the neutrino mixing matrix $V = ({\bf 1} -
\eta)U^{}_0$, which is certainly more non-trivial in this
CP-violating inverse seesaw scenario. By using the formula of
$\eta$ given in Eq. (A7) and the expression of $(M^{}_{\rm
R})^{-1}$ shown in Eq. (25), we obtain
\begin{eqnarray}
\eta^{}_{ee} & \approx & \frac{v^2}{2M^2_0} \left[ \left( 1 + 2
{\rm Re}B^{\prime\prime} \right)^2 + 2 |B^{\prime\prime}|^2
\right] \; , \nonumber \\
\eta^{}_{e\mu} & \approx & \frac{v^2}{2M^2_0} \left[
A^{\prime\prime} {B^{\prime\prime}}^* - \left( 1 + 2 {\rm Re}
B^{\prime\prime} \right) {B^{\prime\prime}}^* - \left( 1 +
A^{\prime\prime} + {B^{\prime\prime}}^* \right)
B^{\prime\prime} \right] \; , \nonumber \\
\eta^{}_{\mu\mu} & \approx & \frac{v^2}{2M^2_0} \left[
{A^{\prime\prime}}^2 + |B^{\prime\prime}|^2 + \left| 1 +
A^{\prime\prime} + B^{\prime\prime} \right|^2 \right] \; , \nonumber \\
\eta^{}_{\mu\tau} & \approx & \frac{v^2}{2M^2_0} \left[
{B^{\prime\prime}}^2 - 2 A^{\prime\prime} \left( 1 +
A^{\prime\prime} + B^{\prime\prime} \right) \right] \; ,
\end{eqnarray}
together with $\eta^{}_{e\tau} = \eta^*_{e\mu}$ and
$\eta^{}_{\tau\tau} = \eta^{}_{\mu\mu}$ for Hermitian $\eta$. Here
we have more free parameters to adjust, such that the values of
$|\eta^{}_{\alpha\beta}|$ (for $\alpha, \beta = e, \mu, \tau$) can
fit their experimental upper bounds given in Eq. (21). Taking
account of $|\eta^{}_{e\mu}| < 3.5 \times 10^{-5}$, we may simply
choose $\eta^{}_{e\mu} \approx 0$ in the calculation of $V$. In
addition, the smallness of both $\theta$ in $U^{}_0$ and that of
$|\eta^{}_{ee}|$, $|\eta^{}_{\mu\mu}|$ and $|\eta^{}_{\mu\tau}|$
allow us to obtain an approximate expression of $V$ as follows:
\begin{eqnarray}
V \approx \left( \matrix{ \displaystyle \frac{2}{\sqrt{6}} \left(
1 - \eta^{}_{ee} \right) & \displaystyle \frac{1}{\sqrt{3}} \left(
1 - \eta^{}_{ee} \right) & \displaystyle -i
\frac{2\theta}{\sqrt{6}} \cr \displaystyle -\frac{1}{\sqrt{6}}
\left( 1 - \eta^{}_{+} \right) - i\frac{\theta}{\sqrt{2}} &
\displaystyle \frac{1}{\sqrt{3}} \left( 1 - \eta^{}_{+} \right) &
\displaystyle \frac{1}{\sqrt{2}} \left( 1 - \eta^{}_{-} \right) +
i\frac{\theta}{\sqrt{6}} \cr \displaystyle -\frac{1}{\sqrt{6}}
\left( 1 - \eta^*_{+} \right) + i \frac{\theta}{\sqrt{2}} &
\displaystyle \frac{1}{\sqrt{3}} \left( 1 - \eta^*_{+} \right) &
\displaystyle -\frac{1}{\sqrt{2}} \left( 1 - \eta^*_{-} \right) +
i\frac{\theta}{\sqrt{6}} \cr} \right) \; ,
\end{eqnarray}
where $\eta^{}_{\pm} \equiv \eta^{}_{\mu\mu} \pm
\eta^{}_{\mu\tau}$ is defined, $\cos\theta \approx 1$ and
$\sin\theta \approx \theta$ are taken, and the terms of ${\cal
O}(\theta \eta^{}_{\alpha\beta})$ are omitted (for $\alpha\beta =
ee, \mu\mu, \mu\tau$). It becomes quite obvious that the
non-unitary corrections to $U^{}_0$ contain a new CP-violating
phase $\arg(\eta^{}_{\mu\tau})$, which is possible to give rise to
an observable effect of CP violation in $\nu^{}_\mu \rightarrow
\nu^{}_\tau$ and $\overline{\nu}^{}_\mu \rightarrow
\overline{\nu}^{}_\tau$ oscillations.

To see the above point more clearly, let us calculate the
following Jarlskog invariants \cite{J} of leptonic CP violation:
\begin{eqnarray}
J^{13}_{\mu\tau} & \equiv & {\rm Im}\left(V^{}_{\mu 1} V^{}_{\tau
3} V^*_{\mu 3} V^*_{\tau 1} \right) \approx \frac{1}{3}
\left( {\rm Im} \eta^{}_{\mu\tau} - \frac{\theta}{\sqrt{3}} \right)
\; , \nonumber \\
J^{23}_{\mu\tau} & \equiv & {\rm Im}\left(V^{}_{\mu 2} V^{}_{\tau
3} V^*_{\mu 3} V^*_{\tau 2} \right) \approx \frac{1}{3} \left( 2
{\rm Im} \eta^{}_{\mu\tau} + \frac{\theta}{\sqrt{3}} \right)  \; .
\end{eqnarray}
It is the sum $J^{13}_{\mu\tau} + J^{23}_{\mu\tau} \approx {\rm
Im}\eta^{}_{\mu\tau}$ that appears in the probabilities of
$\nu^{}_\mu \rightarrow \nu^{}_\tau$ and $\overline{\nu}^{}_\mu
\rightarrow \overline{\nu}^{}_\tau$ oscillations \cite{Xing08},
and thus we have
\begin{eqnarray}
P(\nu^{}_\mu \rightarrow \nu^{}_\tau) & \approx & \sin^2
\frac{\Delta^{}_{32}}{2} ~ + ~ 2 {\rm Im}\eta^{}_{\mu\tau}
\sin\Delta^{}_{32} \;
, \nonumber \\
P(\overline{\nu}^{}_\mu \rightarrow \overline{\nu}^{}_\tau) &
\approx & \sin^2 \frac{\Delta^{}_{32}}{2} ~ - ~ 2 {\rm
Im}\eta^{}_{\mu\tau} \sin\Delta^{}_{32} \; ,
\end{eqnarray}
where $\Delta^{}_{32} \equiv \Delta m^2_{32} L/(2E)$ with $E$
being the neutrino beam energy and $L$ being the baseline length.
In obtaining Eq. (36), we have neglected those small non-unitary
but CP-conserving effects, including the ``zero-distance" effect
\cite{Antusch}. Considering $|{\rm Im}\eta^{}_{\mu\tau}| \leq
|\eta^{}_{\mu\tau}| < 5.0 \times 10^{-3}$ as given in Eq. (21), we
find that it is possible to have non-unitary CP violation at the
percent level in a medium-baseline experiment of $\nu^{}_\mu
\rightarrow \nu^{}_\tau$ and $\overline{\nu}^{}_\mu \rightarrow
\overline{\nu}^{}_\tau$ oscillations (see Ref. \cite{Yasuda} for
more detailed and model-independent discussions).

\section{Summary}

The origin of small masses of three active neutrinos, together
with their unexpectedly large mixing angles, is a big puzzle in
particle physics. Although a lot of attempts have been made in the
past decade to solve this flavor problem, new ideas are eagerly
wanted in the upcoming LHC era to achieve a balance between
theoretical naturalness and experimental testability of the
mechanisms of neutrino mass generation and flavor mixing. In the
present work we have combined the inverse seesaw mechanism with
the FL symmetry to fix the flavor textures of neutrino mass
matrices at the TeV scale, such that both the neutrino mass
spectrum and the neutrino mixing pattern can be calculated. To be
explicit, we have applied the FL symmetry to the gauge-singlet
neutrino sector and shown that it forces one pair of the
gauge-singlet neutrinos to be massless, leading to a simplified
but viable version of the inverse seesaw mechanism which has two
pairs of massive gauge-singlet neutrinos and allows one active
neutrino to be massless. Taking account of the exact $\mu$-$\tau$
permutation symmetry, we have proposed a very simple FL symmetry
breaking ansatz in the gauge-singlet neutrino sector and then
obtained the tri-bimaximal neutrino mixing pattern by means of the
inverse seesaw relation. We find that non-unitary corrections to
this interesting neutrino mixing matrix are possible to reach the
percent level at the TeV scale. We have also demonstrated that a
soft breaking of $\mu$-$\tau$ symmetry can easily accommodate
leptonic CP violation in such an inverse seesaw scenario.

We remark that it is technically more natural to make the inverse
seesaw mechanism work at the TeV scale, although it contains much
more degrees of freedom than the canonical (type-I) seesaw
mechanism. We also remark that the FL symmetry discussed in this
work, like some other flavor symmetries discussed in the
literature, may serve as a phenomenological organizing principle
for studying the flavor textures of neutrino mass matrices. How to
break this symmetry is certainly an open question and involves a
lot of arbitrariness, but our ansatz shows that there exist one or
more simple symmetry breaking patterns which allow us to predict
(or at least to understand) small neutrino masses and large
neutrino mixing angles. As for the TeV-scale inverse seesaw
picture under consideration, it is even possible to have
observable effects, induced by the non-unitarity of the effective
mixing matrix of three active neutrinos, in a delicate medium- or
long-baseline neutrino oscillation experiment.

It is finally worth emphasizing that testing the unitarity of the
light Majorana neutrino mixing matrix in neutrino oscillations and
searching for the signatures of heavy Majorana (or pseudo-Dirac)
neutrinos at the LHC can be complementary to each other, both
qualitatively and quantitatively, in order to deeply understand
the intrinsic properties of Majorana particles \cite{ICHEP08}. In
spite of many challenges on the road ahead, we optimistically
expect that some experimental breakthrough in this aspect will
pave the way towards the true theory of neutrino mass generation,
flavor mixing and CP violation.

\begin{acknowledgments}
One of us (Z.Z.X.) is deeply indebted to K.K. Phua for his warm
hospitality at the IAS of NTU. He is also grateful to S. Luo and
S. Zhou for some useful discussions. This work is supported in
part by the National University of Singapore (academic research
grant No. WBS: R-144-000-178-112) and by the National Natural
Science Foundation of China (under grant No. 10425522 and No.
10875131).
\end{acknowledgments}

\newpage

\appendix

\section{Weak charged-current interactions}

The standard charged-current interactions between three active
neutrinos and three charged leptons are given by
\begin{eqnarray}
-{\cal L}^{}_{\rm cc} \; = \; \frac{g}{\sqrt{2}} \overline{\left(
\matrix{e & \mu & \tau \cr} \right)^{}_{\rm L}} ~ \gamma^\mu
\left( \matrix{\nu^{}_e \cr \nu^{}_\mu \cr \nu^{}_\tau}
\right)^{}_{\rm L} W^-_{\mu} + {\rm h.c.} \;
\end{eqnarray}
in the basis of their flavor eigenstates. Without loss of
generality, we choose the basis in which the flavor eigenstates of
three charged leptons are identified with their mass eigenstates.
We proceed to diagonalize the $9\times 9$ neutrino mass matrix
$\cal M$ in Eq. (3) so as to reexpress ${\cal L}^{}_{\rm cc}$ in
terms of the mass eigenstates of both charged leptons and
neutrinos. Because $\cal M$ is symmetric, it can be diagonalized
by the following unitary transformation:
\begin{eqnarray}
\left( \matrix{V^{}_{3\times 3} & R^{}_{6\times 3} \cr
S^{}_{3\times 6} & U^{}_{6\times 6} \cr} \right)^\dagger {\cal M}
\left( \matrix{V^{}_{3\times 3} & R^{}_{6\times 3} \cr
S^{}_{3\times 6} & U^{}_{6\times 6} \cr} \right)^* = \left(
\matrix{ (\widehat{M}^{}_\nu)^{}_{3\times 3} & {\bf 0} \cr {\bf 0}
& (\widehat{M}^{}_N)^{}_{6\times 6} \cr} \right) \; ,
\end{eqnarray}
in which $\widehat{M}^{}_\nu $ and $\widehat{M}^{}_N$ denote the
diagonal mass matrices of light and heavy neutrinos, respectively,
in the inverse seesaw mechanism. After this diagonalization, the
neutrino flavor eigenstates $\nu^{}_\alpha$ (for $\alpha =e, \mu,
\tau$) can be expressed in terms of the neutrino mass eigenstates
$\nu^{}_i$, $N^{}_i$ and $S^{}_i$ (for $i=1,2,3$) as
\begin{eqnarray}
\nu^{}_{\alpha \rm L} = V \nu^{}_{i \rm L} + R \left( \matrix{
N^{}_i \cr S^{}_i \cr} \right)^{}_{\rm L} \; .
\end{eqnarray}
Substituting this equation into Eq. (A1), we immediately arrive at
\begin{eqnarray}
-{\cal L}^{}_{\rm cc} = \frac{g}{\sqrt{2}}
\overline{\left(\matrix{e & \mu & \tau}\right)^{}_{\rm L}}
~\gamma^\mu \left[ V \left( \matrix{\nu^{}_1 \cr \nu^{}_2 \cr
\nu^{}_3} \right)^{}_{\rm L} + R \left( \matrix{N^{}_1 \cr N^{}_2
\cr N^{}_3 \cr S^{}_1 \cr S^{}_2 \cr S^{}_3 \cr} \right)^{}_{\rm
L} \right] W^-_\mu + {\rm h.c.} \;
\end{eqnarray}
in the basis of mass eigenstates. As $V$ and $R$ belong to the
same unitary transformation done in Eq. (A2), they satisfy the
normalization condition $VV^\dagger + RR^\dagger = {\bf 1}$ and
the exact seesaw relation $V\widehat{M}^{}_\nu V^T + R
\widehat{M}^{}_{N} R^T = {\bf 0}$. Hence $V$ itself must be
non-unitary, and its deviation from unitarity is just measured by
$R$, which actually determines the collider signatures of heavy
Majorana neutrinos at the LHC \cite{Zhou09}. A global analysis of
current experimental data shows that the strength of unitarity
violation of $V$ can at most reach the percent level
\cite{Antusch}.

Let us denote the slight deviation of $V$ from a unitary matrix
$V^{}_0$ as follows: $V = ({\bf 1} - \eta) V^{}_0$ with $\eta
\approx RR^\dagger/2$ being a Hermitian matrix. Then one may
simply follow Ref. \cite{XZZ09} to derive the approximate inverse
seesaw formula given in Eq. (4). Considering $M^{}_{\rm R} \gg
M^{}_{\rm D} \gg M^{}_\mu$ and neglecting small non-unitary
effects (i.e., $V \approx V^{}_0$), we obtain
\begin{eqnarray}
M^{}_\nu \equiv V^{}_0 \widehat{M}^{}_\nu V^T_0 \approx - \left(
\matrix{M^{}_{\rm D} & {\bf 0} \cr} \right) \left( \matrix{ {\bf
0} & M^{}_{\rm R} \cr M^T_{\rm R} & M^{}_\mu \cr} \right)^{-1}
\left( \matrix{ M^T_{\rm D} \cr {\bf 0} \cr} \right) = M^{}_{\rm
D} (M^T_{\rm R})^{-1} M^{}_\mu (M^{}_{\rm R})^{-1} M^T_{\rm D} \;
.
\end{eqnarray}
In this excellent approximation, $V^{}_0$ is actually defined to
be the unitary transformation used to diagonalize the effective
$3\times 3$ mass matrix of three light Majorana neutrinos (i.e.,
$M^{}_\nu$). Because of
\begin{eqnarray}
R \approx \left( \matrix{M^{}_{\rm D} & {\bf 0} \cr} \right)
\left( \matrix{ {\bf 0} & M^{}_{\rm R} \cr M^T_{\rm R} & M^{}_\mu
\cr} \right)^{-1} U \; ,
\end{eqnarray}
where $U$ is approximately unitary to the same degree of accuracy
which assures  Eq. (A5) to hold, we finally arrive at
\begin{eqnarray}
\eta \approx \frac{1}{2} RR^\dagger \approx \frac{1}{2} M^{}_{\rm
D} (M^*_{\rm R} M^T_{\rm R})^{-1} M^\dagger_{\rm D}
\end{eqnarray}
by taking account of $M^{}_{\rm R} \gg M^{}_{\rm D} \gg M^{}_\mu$.
This result allows us to estimate the non-unitary effects in $V$,
once the textures of $M^{}_{\rm D}$ and $M^{}_{\rm R}$ are
specified in an inverse seesaw model.

Note again that we have taken the natural hierarchy $M^{}_{\rm R}
\gg M^{}_{\rm D} \gg M^{}_\mu$ in deriving $M^{}_\nu$ in Eq. (A5)
and calculating $\eta$ in Eq. (A7). In fact, it is mathematically
unnecessary to assume $M^{}_{\rm R} \gg M^{}_\mu$ because the
validity of Eq. (A5) is independent of the relative magnitude of
$M^{}_{\rm R}$ and $M^{}_\mu$. In other words, Eq. (A5) is also
valid if $M^{}_\mu \gg M^{}_{\rm D}$ holds, no matter whether
$M^{}_{\rm R}$ is much larger or smaller than $M^{}_\mu$. But we
reiterate that $M^{}_{\rm R} \gg M^{}_{\rm D} \gg M^{}_\mu$ is a
physical condition of the inverse seesaw mechanism, in which the
smallness of $M^{}_\nu$ is naturally attributed to both the
smallness of $M^{}_\mu$ and the smallness of $M^{}_{\rm
D}/M^{}_{\rm R}$.

\section{Algebraic properties of the FL texture}

We refer to the following form of a mass matrix as the
Friedberg-Lee (FL) texture:
\begin{eqnarray}
{\cal F} = f \left[ \left( \matrix{ 1 & 0 & 0 \cr 0 & 1 & 0 \cr 0
& 0 & 1 \cr} \right) + \left( \matrix{ y + z & -y & -z \cr -y &
x+y & -x \cr -z & -x & x+z \cr}\right) \right] \; ,
\end{eqnarray}
where $f$, $x$, $y$ and $z$ are in general complex parameters.
Given ${\rm Det}{\cal F} \neq 0$, it is easy to show that the
inverse matrix of $\cal F$ takes the same FL texture:
\begin{eqnarray}
{\cal F}^{-1} = \frac{1}{f} \left[ \left( \matrix{ 1 & 0 & 0 \cr 0
& 1 & 0 \cr 0 & 0 & 1 \cr} \right) + \left( \matrix{ y^\prime +
z^\prime & -y^\prime & -z^\prime \cr -y^\prime & x^\prime
+y^\prime & -x^\prime \cr -z^\prime & -x^\prime & x^\prime
+z^\prime \cr}\right) \right] \; ,
\end{eqnarray}
where
\begin{eqnarray}
x^\prime & = & - \frac{x + (xy + yz + zx)}{1 + 2 (x+y+z) + 3 (xy +
yz + zx)} \; , \nonumber \\
y^\prime & = & - \frac{y + (xy + yz + zx)}{1 + 2 (x+y+z) + 3 (xy +
yz + zx)} \; , \nonumber \\
z^\prime & = & - \frac{z + (xy + yz + zx)}{1 + 2 (x+y+z) + 3 (xy +
yz + zx)} \; .
\end{eqnarray}
Now let us consider another mass matrix of the FL texture:
\begin{eqnarray}
{\cal D} = d \left[ \left( \matrix{ 1 & 0 & 0 \cr 0 & 1 & 0 \cr 0
& 0 & 1 \cr} \right) + \left( \matrix{ \beta + \gamma & -\beta &
-\gamma \cr -\beta & \alpha + \beta & -\alpha \cr -\gamma &
-\alpha & \alpha + \gamma \cr}\right) \right] \; ,
\end{eqnarray}
where $d$, $\alpha$, $\beta$ and $\gamma$ are in general complex
parameters. A lengthy but straightforward calculation shows that
the seesaw-like mass matrix ${\cal E} \equiv {\cal D}{\cal
F}^{-1}{\cal D}^T$ takes the same texture as ${\cal D}$ and $\cal
F$ do:
\begin{eqnarray}
{\cal E} = \frac{d^2}{f} \left[ \left( \matrix{ 1 & 0 & 0 \cr 0 &
1 & 0 \cr 0 & 0 & 1 \cr} \right) + \left( \matrix{ \widehat\beta +
\widehat\gamma & -\widehat\beta & -\widehat\gamma \cr
-\widehat\beta & \widehat\alpha + \widehat\beta & -\widehat\alpha
\cr -\widehat\gamma & -\widehat\alpha & \widehat\alpha +
\widehat\gamma \cr}\right) \right] \; ,
\end{eqnarray}
where
\begin{eqnarray}
\widehat\alpha & = & +\alpha \left[ \left( 1 + \alpha + \beta
\right) \left( 1 + x^\prime + y^\prime \right) + \alpha x^\prime +
\beta y^\prime \right] - \gamma \left[ \left( 1 + \alpha + \beta
\right) y^\prime + \beta \left( 1 + y^\prime + z^\prime \right) -
\alpha z^\prime \right] \nonumber \\
&& + \left( 1 + \alpha + \gamma \right) \left[ \left( 1 + \alpha +
\beta \right) x^\prime + \alpha \left( 1 + x^\prime + z^\prime
\right) - \beta z^\prime \right] \; , \nonumber \\
\widehat\beta & = & -\alpha \left[ \left( 1 + \beta + \gamma
\right) z^\prime + \gamma \left( 1 + x^\prime + z^\prime \right) -
\beta x^\prime \right] + \beta \left[ \left(1 + \beta +
\gamma\right) \left(1 + y^\prime + z^\prime\right) + \beta
y^\prime + \gamma z^\prime \right] \nonumber \\
&& + \left( 1 + \alpha + \beta \right) \left[ \left( 1 + \beta +
\gamma \right) y^\prime + \beta \left( 1 + x^\prime + y^\prime
\right) - \gamma x^\prime \right] \; , \nonumber \\
\widehat\gamma & = & -\alpha \left[ \left( 1 + \beta + \gamma
\right) y^\prime + \beta \left( 1 + x^\prime + y^\prime \right) -
\gamma x^\prime \right] + \gamma \left[ \left( 1 + \beta + \gamma
\right) \left( 1 + y^\prime + z^\prime \right) + \beta y^\prime
+ \gamma z^\prime \right] \nonumber \\
&& + \left( 1 + \alpha + \gamma \right) \left[ \left( 1 + \beta +
\gamma \right) z^\prime + \gamma \left( 1 + x^\prime + z^\prime
\right) - \beta x^\prime \right] \; .
\end{eqnarray}
This interesting seesaw-invariant property is a salient feature of
the FL texture.


\begin{thebibliography}{99}
\bibitem{PDG08} Particle Data Group, C. Amsler {\it et al.},
Phys. Lett. B {\bf 667}, 1 (2008).

\bibitem{SS1} H. Fritzsch, M. Gell-Mann, and P. Minkowski, Phys.
Lett. B {\bf 59}, 256 (1975); T.P. Cheng, Phys. Rev. D {\bf 14},
1367 (1976); P. Minkowski, Phys. Lett. B {\bf 67}, 421 (1977); T.
Yanagida, in {\it Proceedings of the Workshop on Unified Theory
and the Baryon Number of the Universe}, edited by O. Sawada and A.
Sugamoto (KEK, Tsukuba, 1979), p. 95; M. Gell-Mann, P. Ramond, and
R. Slansky, in {\it Supergravity}, edited by P. van Nieuwenhuizen
and D. Freedman (North Holland, Amsterdam, 1979), p. 315; S.L.
Glashow, in {\it Quarks and Leptons}, edited by M. L$\acute{\rm
e}$vy {\it et al.} (Plenum, New York, 1980), p. 707; R.N.
Mohapatra and G. Senjanovic, Phys. Rev. Lett. {\bf 44}, 912
(1980).

\bibitem{Xing09} For a recent review, see: Z.Z. Xing, arXiv:0905.3903;
and references therein.

\bibitem{KS} J. Kersten and A.Yu. Smirnov, Phys. Rev. D
{\bf 76}, 073005 (2007).

\bibitem{MV} D. Wyler and L. Wolfenstein, Nucl. Phys. B {\bf 218}, 205 (1983);
R.N. Mohapatra and J.W.F. Valle, Phys. Rev. D {\bf 34}, 1642
(1986); E. Ma, Phys. Lett. B {\bf 191}, 287 (1987).

\bibitem{Zhou09} Z.Z. Xing and S. Zhou, Phys. Lett. B {\bf 679}, 249 (2009).

\bibitem{Fogli} See, e.g., G.L. Fogli {\it et al.}, Phys. Rev. D {\bf 78},
033010 (2008).

\bibitem{Valle} M. Hirsch, S. Morisi, and J.W.F. Valle,
arXiv:0905.3056.

\bibitem{TBM} P.F. Harrison, D.H. Perkins, and W.G. Scott, Phys.
Lett. B {\bf 530}, 167 (2002); Z.Z. Xing, Phys. Lett. B {\bf 533},
85 (2002); P.F. Harrison and W.G. Scott, Phys. Lett. B {\bf 535},
163 (2002); X.G. He and A. Zee, Phys. Lett. B {\bf 560}, 87
(2003).

\bibitem{FL} R. Friedberg and T.D. Lee, High Energy Phys. Nucl.
Phys. {\bf 30}, 591 (2006);  Annals Phys. {\bf 323}, 1087 (2008);
Annals Phys. {\bf 323}, 1677 (2008); T.D. Lee, Nucl. Phys. A {\bf
805}, 54 (2008).

\bibitem{Hooft} G. 't Hooft, in {\it Proceedings of 1979 Carg$\grave{e}$se
Institute on Recent Developments in Gauge Theories}, edited by G.
't Hooft {\it et al.} (Plenum Press, New York, 1980), p. 135.

\bibitem{J08} C. Jarlskog, Phys. Rev. D {\bf 77}, 073002 (2008);
arXiv:0806.2206.

\bibitem{Zhang} M. Malinsky, T. Ohlsson, Z.Z. Xing, and H. Zhang,
Phys. Lett. B {\bf 679}, 242 (2009).

\bibitem{XZZ} Z.Z. Xing, H. Zhang, and S. Zhou, Phys. Lett. B {\bf
641}, 189 (2006).

\bibitem{LX} S. Luo and Z.Z. Xing, Phys. Lett. B {\bf 646}, 242
(2007); Z.Z. Xing, Int. J. Mod. Phys. E {\bf 16}, 1361 (2007); W.
Chao, S. Luo, and Z.Z. Xing, Phys. Lett. B {\bf 659}, 281 (2008);
C.S. Huang, T.J. Li, W. Liao, and S.H. Zhu, Phys. Rev. D {\bf 78},
013005 (2008); T. Araki and R. Takahashi, arXiv:0811.0905; T.
Araki and C.Q. Geng, arXiv:0906.1903.

\bibitem{LXL} S. Luo, Z.Z. Xing, and X. Li, Phys. Rev. D {\bf 78},
117301 (2008).

\bibitem{Antusch} S. Antusch {\it et al.}, JHEP {\bf
0610}, 084 (2006); A. Abada {\it et al.}, JHEP {\bf 0712}, 061
(2007).

\bibitem{FX01} H. Fritzsch and Z.Z. Xing, Phys. Lett. B {\bf 517},
363 (2001).

\bibitem{J} C. Jarlskog, Phys. Rev. Lett. {\bf 55}, 1039 (1985);
D.D. Wu, Phys. Rev. D {\bf 33}, 860 (1986).

\bibitem{Xing08} Z.Z. Xing, Phys. Lett. B {\bf 660}, 515
(2008).

\bibitem{Yasuda} E. Fernandez-Martinez, M.B. Gavela, J.
L$\rm\acute{o}$pez-Pav$\rm\acute{o}$n, and O. Yasuda, Phys. Lett.
B {\bf 649}, 427 (2007); Z.Z. Xing, Phys. Lett. B {\bf 660}, 515
(2008); S. Luo, Phys. Rev. D {\bf 78}, 016006 (2008); S. Goswami
and T. Ota, Phys. Rev. D {\bf 78}, 033012 (2008); G. Altarelli and
D. Meloni, Nucl. Phys. B {\bf 809}, 158 (2009); S. Antusch, M.
Blennow, E. Fernandez-Martinez, and J.
L$\rm\acute{o}$pez-Pav$\rm\acute{o}$n, Phys. Rev. D {\bf 80},
033002 (2009).

\bibitem{ICHEP08} For a recent review, see:
Z.Z. Xing, Int. J. Mod. Phys. A {\bf 23}, 4255 (2008).

\bibitem{XZZ09} Z.Z. Xing, arXiv:0902.2469; Phys. Lett. B {\bf 679}, 255 (2009).

\end{thebibliography}
\end{document}